\documentclass[%
reprint,
superscriptaddress,
showpacs,preprintnumbers,
amsmath,amssymb,
aps,
pra,
longbibliography
]{revtex4-2}
\usepackage{color}
\usepackage{graphicx}
\usepackage{bm}
\usepackage{mathrsfs}
\usepackage{dsfont}
\usepackage{ulem}
\usepackage[colorlinks,allcolors=blue]{hyperref}

\makeatletter

\newcommand{\Rmnum}[1]{\expandafter\@slowromancap\romannumeral #1@}

\makeatother
\def\bra#1{\langle #1|}
\def\ket#1{ |#1 \rangle}

\begin{document}
\title{Quantum state transfer between two photons with polarization and orbital angular momentum via quantum teleportation technology}

\author{Shihao Ru}
 \affiliation{Shaanxi Key Laboratory of Quantum Information and Quantum Optoelectronic Devices, School of Physics of Xi'an Jiaotong University, Xi'an 710049, China}
 \author{Min An}
\affiliation{Shaanxi Key Laboratory of Quantum Information and Quantum Optoelectronic Devices, School of Physics of Xi'an Jiaotong University, Xi'an 710049, China}%
\author{Yu Yang}
\affiliation{Shaanxi Key Laboratory of Quantum Information and Quantum Optoelectronic Devices, School of Physics of Xi'an Jiaotong University, Xi'an 710049, China}%
\author{Rui Qu}
\affiliation{Shaanxi Key Laboratory of Quantum Information and Quantum Optoelectronic Devices, School of Physics of Xi'an Jiaotong University, Xi'an 710049, China}%
\author{Feiran Wang}
\affiliation{Shaanxi Key Laboratory of Quantum Information and Quantum Optoelectronic Devices, School of Physics of Xi'an Jiaotong University, Xi'an 710049, China}%
\affiliation{School of Science of Xi’an Polytechnic University, Xi'an 710048, China}%
\author{Yunlong Wang}%
\email{yunlong.wang@mail.xjtu.edu.cn}
\affiliation{Shaanxi Key Laboratory of Quantum Information and Quantum Optoelectronic Devices, School of Physics of Xi'an Jiaotong University, Xi'an 710049, China}%
\author{Pei Zhang}
\affiliation{Shaanxi Key Laboratory of Quantum Information and Quantum Optoelectronic Devices, School of Physics of Xi'an Jiaotong University, Xi'an 710049, China}%
\author{Fuli Li}
\affiliation{Shaanxi Key Laboratory of Quantum Information and Quantum Optoelectronic Devices, School of Physics of Xi'an Jiaotong University, Xi'an 710049, China}%
\date{\today}

\begin{abstract}
Quantum teleportation is a useful quantum information technology to transmit quantum states between different degrees of freedom. We here report a quantum state transfer experiment in the linear optical system, transferring a single photon state in the polarization degree of freedom (DoF) to another photon in the orbital angular momentum (OAM) quantum state via a biphoton OAM entangled channel. Our experimental method is based on quantum teleportation technology. The differences between ours and the original teleportation scheme is that the transfer state is known in ours, and our method is for different particles with different DoFs while the original one is for different particles with same DoF. Besides, our present experiment is implemented with a high Bell-efficiency since each of the four hybrid-entangled Bell states can be discriminated. We use six states of poles of the Bloch sphere to test our experiment, and the fidelity of the quantum state transfer is $91.8\pm1.3\%$.
\end{abstract}

\maketitle
\section{Introduction}
\label{sec:introduction}
Quantum entanglement is one of the most important resource for quantum information \cite{nielsen_chuang_2010,RMPentanglement,erhard2020advances}. It plays a central role in many applications, i.e., quantum precision measurements \cite{giovannetti2011advances}, fault tolerant quantum computation \cite{GottesmanFTQC,2020TFTQC}, quantum network \cite{kimble2008quantum,simon2017towards,wehner2018quantum}, and many meaningful quantum technologies.
Quantum teleportation\cite{pirandola2015advances,zeilinger2018quantum}, as a celebrated quantum technology involving the entanglement, aims to transfer the quantum state between the different physical carriers artfully.
We here consider a different scenario where the polarization DoF on Alice side is convenient for quantum information processing, but OAM DoF on Bob side is advanced.
A quantum task is thus raised to transmit a polarization encoding state from Alice to an OAM encoding qubit on Bob's side. This quantum task can be solved by virtue of a teleportation-based method.

Quantum teleportation is a fundamental protocol in quantum information with no any classical analog \cite{pirandola2015advances,zeilinger2018quantum}. It allows for the simulation of an ideal quantum channel by exploiting entanglement and classical communication.
One of the most important features of quantum teleportation is that a quantum state can be transmitted faithfully, even though the state itself is completely unknown to the sender and the receiver as well.
In the past decades, quantum teleportation has been extensively developed on many physical platforms, such as linear optics \cite{bouwmeester1997experimental,wang2015quantum}, trapped ions \cite{barrett2004deterministic,riebe2004deterministic,nolleke2013efficient}, superconducting circuits \cite{baur2012benchmarking,steffen2013deterministic}, even  some high-dimensional systems \cite{zhang2019quantum,luo_tele,xiaomin_tele}, hybrid-DoF discrete systems \cite{PhysRevA.87.022326,PhysRevA.83.060301,PhysRevA.80.063831}, and hybrid-entangled systems between continuous and discrete variables \cite{sychev2018entanglement,PhysRevLett.118.160501,takeda2013deterministic,2017hybridstmo,huo2018deterministic}.

Besides, realizing a complete Bell-state measurement (BSM) is still a major issue for quantum teleportation, which directly affects the efficiency of the quantum state transfer.
Recalling that the simplest BSM scenario in the linear optics with the necessary single-photon detectors only allows at most two of the four Bell states to be distinguished, and the corresponding Bell-efficiency is constrained to be 50\% \cite{lutkenhaus1999bell,vaidman1999methods}.
For improving the Bell-efficiency as high as possible, many schemes have been presented in the past years.
For example, Ref. \cite{2017williams} has employed the extra time DoF to achieve this mission and its Bell efficiency could be promoted into 100\% in principle.
Ref. \cite{zhang2019arbitrary} has proposed an arbitrary bipartite high-dimensional BSM configuration with the auxiliary entanglement.
Ref. \cite{gao2020universal} has exploited the technologies of hyper-entanglement, which divided the 16 hyper-entangled Bell states into 14 groups.
This theoretical possibility clearly implies an overhead of quantum resources that is a nontrivial experimental challenge. Solving this problem is an active area of research \cite{2017williams,zhang2019arbitrary,gao2020universal}.

In this paper, our work mainly focuses on the quantum state transfer on the photon (also named ``flying qubit") that is the most ideal transmission and processing carrier of information \cite{E91}.
Particularly, the polarization DoF of the photon is frequently employed for coding a variety of information \cite{1996densecoding,1997tele}.
Apart from the polarization DoF, the photonic transverse spatial mode DoF \cite{erhard2018twisted,forbes2019quantum,wang2020vectorial} has attached wide attention recently since its broad prospects in the area of quantum communication.
By way of illustrations, A. Zeilinger et al. \cite{resch2005distributing} has distributed the entangled photons directly through the atmosphere to a receiver station 7.8 km away over the city of Vienna, and they \cite{krenn2015twisted,krenn2014communication} also verified quantum entanglement of photon pairs with spatial modes over a turbulent, real-world link of 3 km across Vienna.
A. Forbes et al. has demonstrated the feasibility of transferring the multi-dimensional OAM entangled states over $250$ m through the single-mode fiber \cite{liu2020multidimensional}.
Y. Cao et al. has reported the three-dimensional OAM entanglement distribution via a 1-km-long fewer-mode optical fiber by using an actively stabilizing phase precompensation technique \cite{cao2020distribution}.

In this work, we implement the quantum state transfer of different photons from the polarization space to the OAM space.
This paper is organized as follows. In Sec. \ref{sec:proposal}, the experimental proposal is introduced, which intriguingly combines the conventional teleportation protocol \cite{1993Bennett} with a new presented hybrid-DoF BSM.
Moreover, we show that the Bell-efficiency of this hybrid-DoF BSM can be theoretically reached 100\%.
In  Sec. \ref{sec:experimental}, the experimental setup and results are detailedly exhibited.
We use six states of poles of the Bloch sphere to test our scheme, and the fidelity of the quantum state transfer being $91.8\pm1.3\%$.
Our work gives a reasonable polarization-OAM quantum interface solution for the quantum state transfer, and contributes a hybrid-entangled BSM approach for achieving a tangible Bell-efficiency enhancement.

\section{Experimental Proposal}
\label{sec:proposal}
\begin{figure}[!t]
	\centering
	\includegraphics[width=0.95\linewidth]{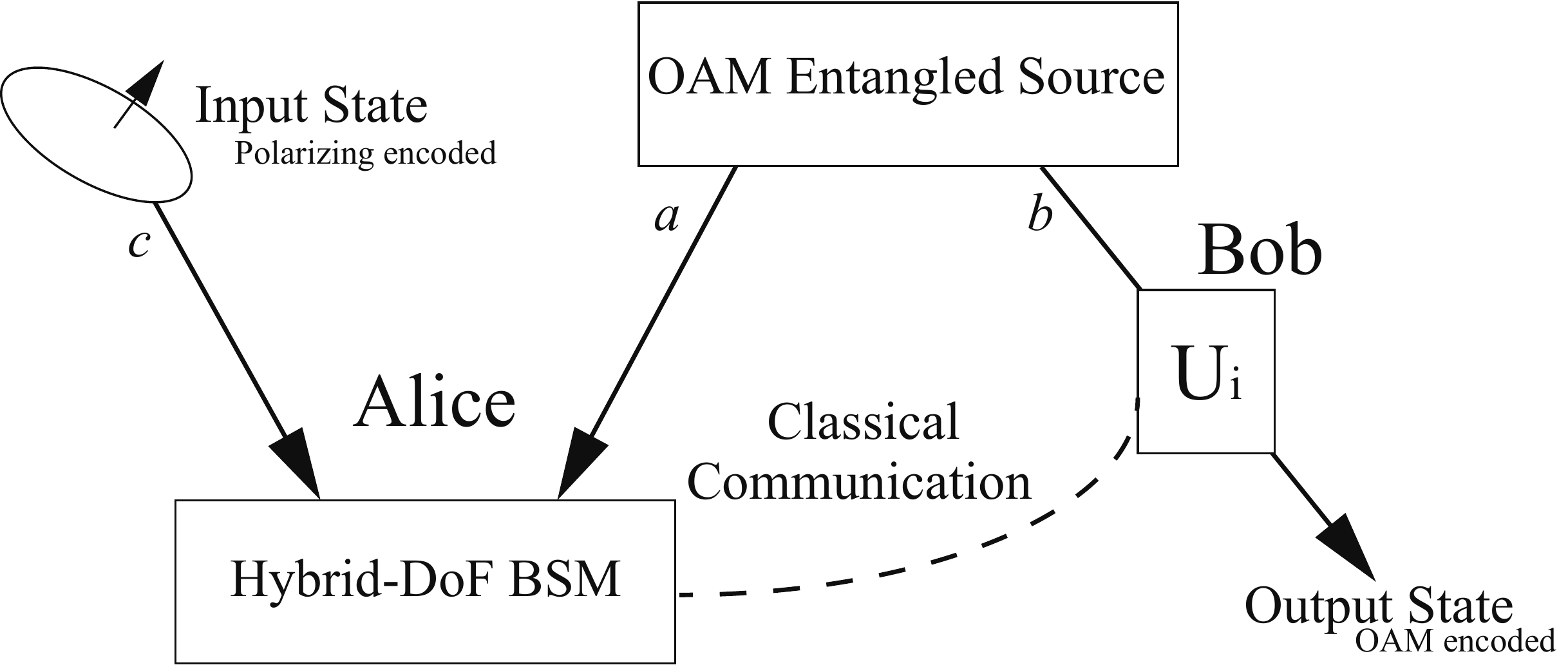}
	\caption{Scheme of quantum state transfer. A polarization input state $c$ at Alice is transmitted to Bob using a shared OAM maximum entangled state $\ket{\phi_-}_{ab}$ and a classical communication channel. Alice performs a Hybrid-DoF Bell state measurement on her systems, $a$ and $c$, and communicates the outcome to Bob. Bob can check his qubit $b$ under the conditional outcome information from Alice.}\label{fig:tele:o}
\end{figure}

Our experimental proposal is shown in FIG. \ref{fig:tele:o}.  We consider two parties --- Alice and Bob --- who share two qubits of OAM DoF, $a$ and $b$. The two qubits are prepared in a maximally OAM entangled state, which is generated by a Hong-Ou-Mandel (HOM) interference \cite{1987HOM,yu2019,bouchard2020two}.
Let us consider that two photons incident on two input ports of a beam splitter (BS), respectively. When arriving at input ports of the BS, the two photons are in the state
\begin{align}
	\ket{\psi_{\mathrm{in}}}_{ab}=\hat{a}_j^{\dagger}\hat{b}_k^{\dagger}\ket{\mathrm{vacumn}}_{ab}=\ket{1;j}_a\ket{1;k}_b,
\end{align}
where $\hat{a}_j^{\dagger}$ and $\hat{b}_k^{\dagger}$ are creation operators of photons in $a$ and $b$ modes of BS, respectively. The subscripts $j$ and $k$ denotes other properties of the photons, which determine how distinguishable they are, such as polarization and transverse spatial modes, and so on.
Specifically, we consider that two photons interfere on a 50:50 BS. If the two photons are in different polarization states, i.e., one is in $\ket{H}$ and another in $\ket{V}$, the corresponding output state is
\begin{align}
\ket{\psi_{\mathrm{out}}}_{ab}&=\frac{1}{2}{\Big(}\ket{2;H,V}_a+\ket{2;H,V}_b\notag\\
&+\ket{1;H}_{a}\ket{1;V}_b-\ket{1;V}_a\ket{1;H}_b{\Big)}.
\end{align}
The first and second terms of the state represent the case where the two photons simultaneously exit out of the same port of the BS (photon bunching), and the last two components describe the situation where the two photons leave out of the different ports (photon anti-bunching), respectively.
If the two photons are in the same polarization state, i.e., $\ket{H}_a\ket{H}_b$, the corresponding output state is
\begin{align}
\ket{\psi_{\mathrm{out}}}_{ab}=\frac{1}{\sqrt{2}}\left(\ket{2;H,H}_a-\ket{2;H,H}_b\right).
\end{align}

The OAM maximally entangled state, $\ket{\phi_-}_{ab}=(\ket{+\ell_0}_a\ket{+\ell_0}_b-\ket{-\ell_0}_a\ket{-\ell_0}_b)/\sqrt{2}$ (known as a Bell pair), can be generated in a similar way.
For convenience, $\ket{\pm\ell_0}$ denote OAM states with different OAM values of $\ell=+1$ and $\ell=-1$, respectively.
When a pair of input photons is in the same OAM state ($\ket{+\ell_0}$), since its spatial mode will be flipped after reflection, the output state is
\begin{align}\label{formula_OAM}
\ket{\psi_{\mathrm{out}}}_{ab}&=\frac{1}{2}{\Big(}\ket{2;+\ell_0,-\ell_0}_a+\ket{2;+\ell_0,-\ell_0}_b\notag\\
&+\ket{1;+\ell_0}_{a}\ket{1;+\ell_0}_b-\ket{1;-\ell_0}_a\ket{1;-\ell_0}_b{\Big)}.
\end{align}
Since the present experimental scheme has a post-selection count coincidence between modes $a$ and $b$, and the first and second components of (\ref{formula_OAM}) is automatically filtered, the state (\ref{formula_OAM}) is just the entangled state $\ket{\phi_-}_{ab}$ required for our experimental scheme.

\begin{figure*}[!hbtp]
	\centering
	\includegraphics[width=0.83\linewidth]{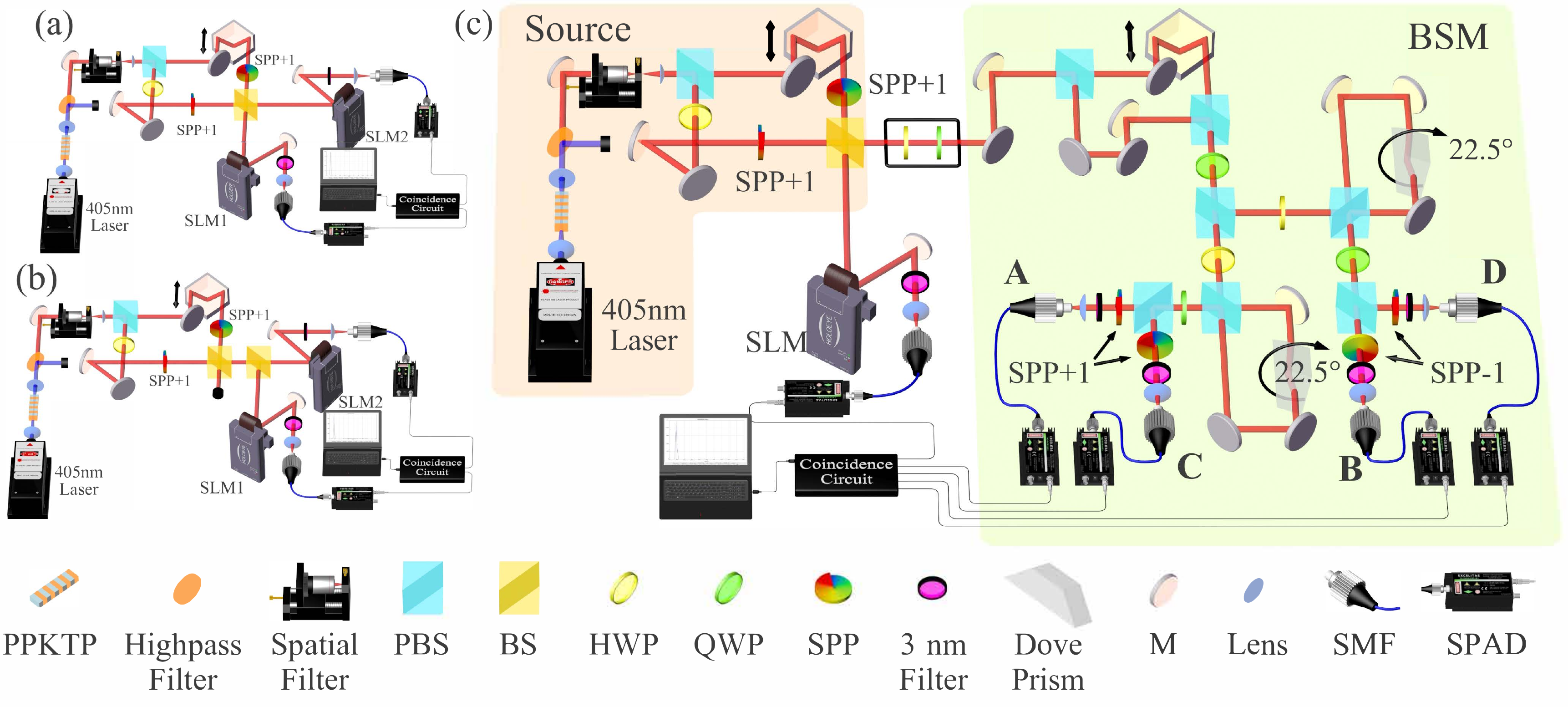}
	\caption{(color online) Experimental Setup (a) and (b) are to measure OAM quantum states in anti-bunching and bunching, respectively. (c) Experimental setup for polarization-OAM quantum state transfer, in which OAM entangled source is located in the pink box, any polarization quantum state of Alice's qubit $c$ can be prepared by the HWP and QWP located in the black box (Between Source and BSM Parts), and Hybrid-DoFs (OAM and polarization) Bell state measurement is located in the green box. PPKTP: poled potassium titanyl phosphate; PBS: polarizing beam splitter; BS: beam splitter; HWP: half wave plate; QWP: quart wave plate; SPP: spiral phase plate; M: mirror; SMF: single mode fiber; SPAD: single-photon avalanche detector.}\label{fig:exp}
\end{figure*}

As shown in FIG. \ref{fig:tele:o},  Alice and Bob share the entangled state $\ket{\phi_-}_{ab}$ meanwhile 
 another qubit $c$ in the polarizing state $\ket{\psi}_c$ is sent to Alice
\begin{equation}
\ket{\psi}_c=\alpha\ket{H}_c+\beta\ket{V}_c,
\end{equation}
where $\alpha^2+\beta^2=1$ and Bob does not know any information of the state.
The whole hybrid-DoF quantum state which Alice and Bob hold can written as
\begin{align}\label{fullstate}
\ket{\Psi}&=\ket{\psi}_c\otimes\ket{\phi_-}_{ab}\notag\\
			    &=(\alpha\ket{H}_c+\beta\ket{V}_c)\notag\\
			    &\otimes\frac{1}{\sqrt{2}}(\ket{+\ell_0}_a\ket{+\ell_0}_b-\ket{-\ell_0}_a\ket{-\ell_0}_b).
\end{align}
We define  the four hybrid-entangled Bell states in OAM and polarization DoFs
\begin{align}\label{Bellstates}
&\ket{\omega_{ca}^{\pm}}=\frac{1}{\sqrt{2}}(\ket{H}_c\ket{+\ell_0}_a\pm\ket{V}_c\ket{-\ell_0}_a),\notag\\
&\ket{\xi_{ca}^{\pm}}=\frac{1}{\sqrt{2}}(\ket{H}_c\ket{-\ell_0}_a\pm\ket{V}_c\ket{+\ell_0}_a).
\end{align}
In terms of these states, Eq.~(\ref{fullstate}) can be rewritten as
\begin{align}
\ket{\Psi}=\frac{1}{2}&{\big[}\ket{\omega_{ca}^+}(\alpha\ket{+\ell_0}_b-\beta\ket{-\ell_0}_b)\notag\\
&+\ket{\omega_{ca}^-}(\alpha\ket{+\ell_0}_b+\beta\ket{-\ell_0}_b)\notag\\
&+\ket{\xi_{ca}^+}(\beta\ket{+\ell_0}_b-\alpha\ket{-\ell_0}_b)\notag\\
&-\ket{\xi_{ca}^-}(\beta\ket{+\ell_0}_b+\alpha\ket{-\ell_0}_b){\big]}.
\end{align}

A vital step for our experimental scheme is to perform perfect BSM of qubits $c$ and $a$, projecting them onto the basis of the four hybrid-entangled Bell states  (\ref{Bellstates}), and discriminating each one of them.
In our scenario, Alice performs a complete BSM, which projects her qubits $a$ and $c$ into one of the four Bell states.
As a result, the state of the input qubit collapses due to measurement meanwhile the qubit $b$ in Bob side is simultaneously projected onto the certain quantum state that is different from the input state only by a unitary transformation. 
In the feed-forward step, Alice communicates the outcome of her measurement through a classical channel with Bob, who then applies the corresponding unitary operation to recover the original input state on the qubit $b$.
Similarly for experimental verification, Bob only needs to perform quantum state distinguish on qubit $c$ after classical communication and observe whether it is consistent.

Note that Alice's input state is assumed to be unknown, otherwise it reduces to remote state preparation.
In typical experiments, the input state is taken to be pure and belonging to a limited alphabet, for example, the six poles of the Bloch sphere.
In the presence of decoherence, the quality of the reconstructed state may be quantified by its fidelity $F\in(0,1]$.
This is the fidelity between Alice's input state and Bob's output state, averaged over all the outcomes of the BSM and input state alphabet.
In the original teleportation scheme, for small values of the fidelity, strategies exist that allow for an imperfect teleportation while making no use of any entangled resource.
For example, Alice may directly measure her input state, thereby sending the results to Bob for him to prepare an output state.
Such a measure–prepare strategy is known as ‘classical teleportation’ and has the maximum fidelity $F_{\mathrm{class}}=2/3$ for an arbitrary input state or, equivalently, an alphabet of mutually unbiased states, such as the six poles of the Bloch sphere.
Thus, the requirement $F>F_{\mathrm{class}}$ is a clear benchmark for ensuring that quantum resources are utilized \cite{Classicaltele}.
\begin{figure}[t]
	\centering
	\includegraphics[width=\linewidth]{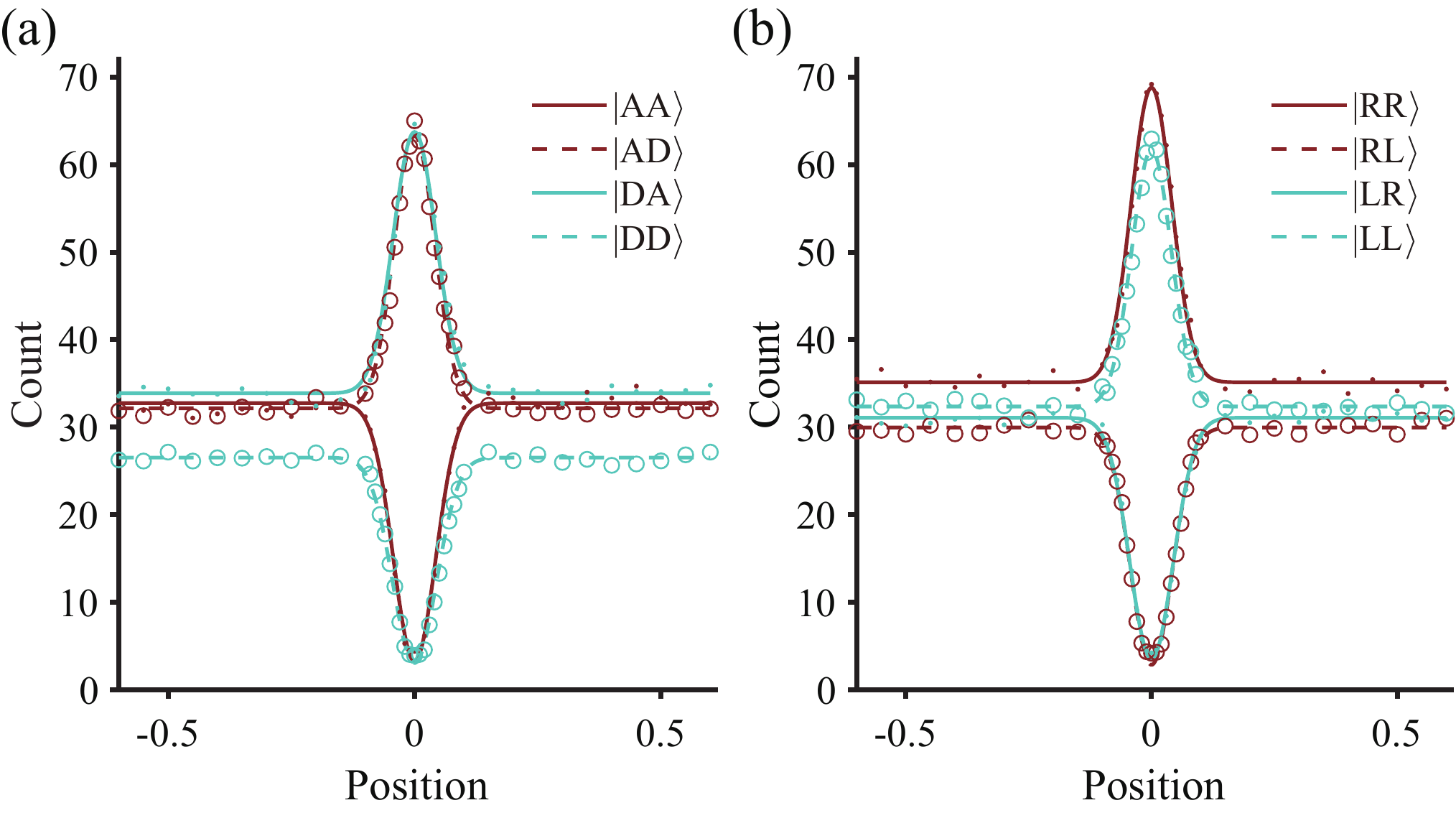}
	\caption{(color online) (a) and (b) are two-photon OAM HOM interference results for Fig.~\ref{fig:exp}(a) on different measurement bases, where $\ket{D}=(\ket{+\ell_0}+\ket{-\ell_0})/\sqrt{2}$, $\ket{A}=(\ket{+\ell_0}-\ket{-\ell_0})/\sqrt{2}$, $\ket{R}=(\ket{+\ell_0}+\mathrm{i}\ket{-\ell_0})/\sqrt{2}$, and $\ket{L}=(\ket{+\ell_0}-\mathrm{i}\ket{-\ell_0})/\sqrt{2}$. In addition, since the fitting curves are close to each other and the error bars are small under the coordinate of this figure, error bars are not drawn for a better presentation.}\label{fig:expdata1}
\end{figure}
\begin{figure*}[t]
	\centering
	\includegraphics[width=0.8\linewidth]{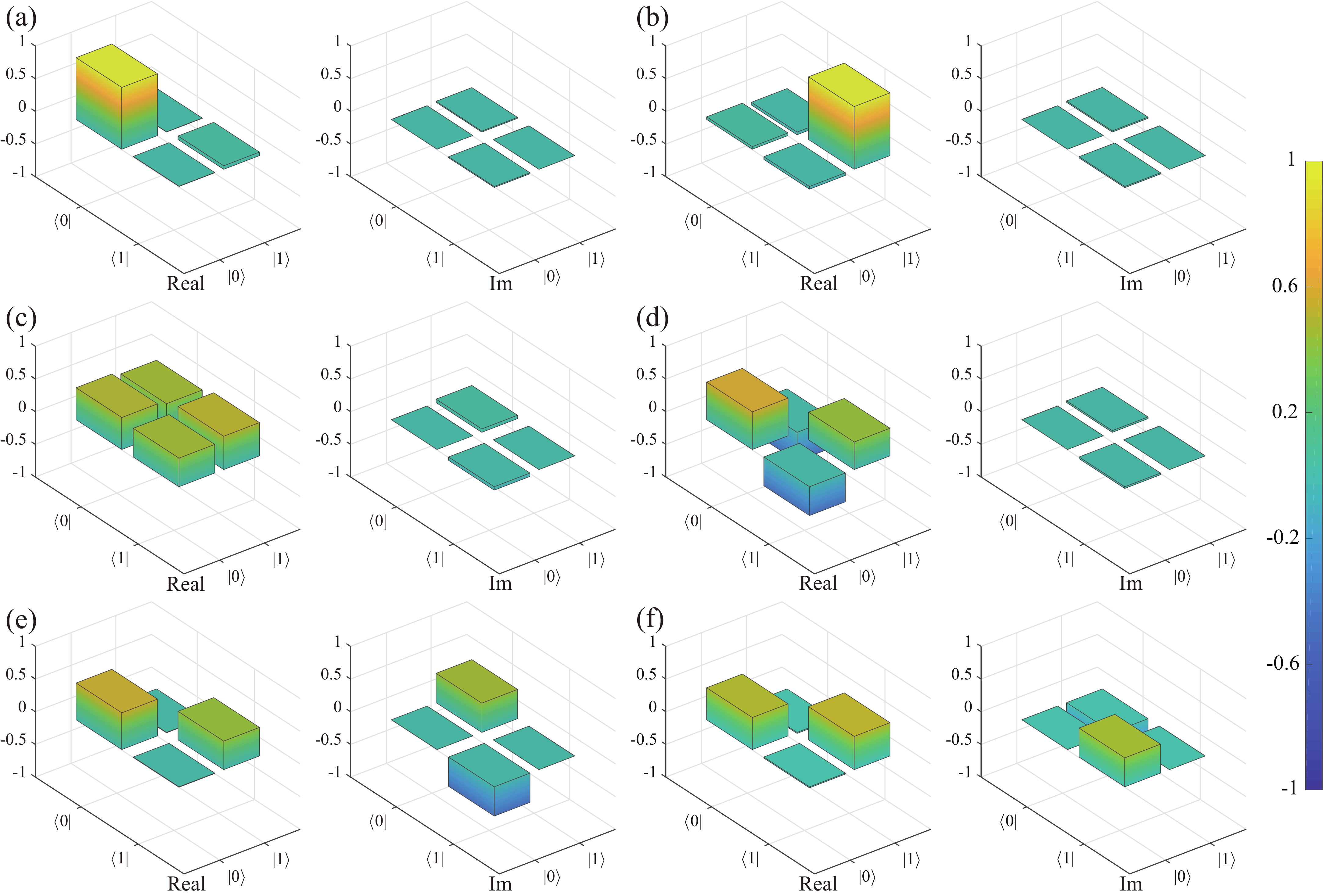}
	\caption{(color online) Tomography of the transmitted states. The input states prepared on qubit $c$ are the six poles of the Bloch sphere, $\ket{0}=\ket{H}$, $\ket{1}=\ket{V}$, $\ket{D}=(\ket{H}+\ket{V})/\sqrt{2}$, $\ket{A}=(\ket{H}-\ket{V})/\sqrt{2}$, $\ket{R}=(\ket{H}+\mathrm{i}\ket{V})/\sqrt{2}$ and $\ket{L}=(\ket{H}-\mathrm{i}\ket{V})/\sqrt{2}$, respectively. Colored bars show the experimental results.}\label{fig:Alltomo}
\end{figure*}

\section{Experimental setup and results}
\label{sec:experimental}
FIG. \ref{fig:exp} shows the experimental setup for polarization-OAM quantum state transfer. A continuous wave laser with center wavelength at 405 nm is used to pump a 2 mm long linear crystal of periodically poled potassium titanyl phosphate (PPKTP), cutting for degenerate type-\Rmnum{2} collinear phase matching which emits correlated photon pairs at 810 nm.
Photons from the source are spatially filtered to the fundamental Gaussian mode using a spatial filter.
The polarizing beam splitter (PBS) is to separate correlated photon pairs.
A half wave plate (HWP) is used to make the polarizations of correlated photon pairs be consistent, and spiral phase plates (SPP, topological charge $+1$) are used to modulate photon pairs in the same OAM state $\ket{+\ell_0}$.

There is a motorized translation stage in the Mach-Zehnder interferometer, which is used to control the time difference of correlated photon pairs to the BS. Its traveling range is 27 mm, and the minimum incremental motion is 0.2 $\mu$m.
As shown by formula (\ref{formula_OAM}), due to the OAM HOM interference of photon pairs at the BS, both the anti-bunching  and bunching  components exist.
Experimentally, we use the setups illustrated in FIG.~\ref{fig:exp}(a) and~\ref{fig:exp}(b) to study the aforementioned two-photon correlation characteristics, respectively.
For detection, a spatial light modulator (SLM) together with a SMF is used to perform any directional projection measurements of OAM modes.  The coincidence counts of two single-photon avalanche detectors are proportional to the detected correlated photon pairs.

We have measured eight kinds of two-photon joint measurements for FIG. \ref{fig:exp}, and its results are shown in FIG.~\ref{fig:expdata1}, in which the position from $-0.6$ mm to $0.6$ mm is the range of movement of the motorized translation stage.
It can be observed that the half width of HOM interference is about 194 $\mu$m as the narrowband of filters used here is 3 nm.
Therefore, the OAM entangled state $\ket{\phi_-}_{ab}$, i.e. anti-bunching part in Eq. (\ref{formula_OAM}), is generated at the origin of the coordinate.
The fidelity between the theoretical state $\ket{\phi_-}_{ab}$ and experimentally prepared state is $92.55\pm1.02\%$.
Similarly for FIG. \ref{fig:exp}(b), we have checked the bunching part in Eq. (\ref{formula_OAM}) and its fidelity is $93.13\pm1.21\%$.

The transmitted qubit here is encoded in polarization DoF and can be prepared in an arbitrary pure state by the HWP and QWP, which are located in the black box.
Now let us turn to BSM, which is a vital step for our experiment.
For multi-DoF systems, a complete BSM can be realized without any auxiliary qubits. As shown in the green part of FIG.~\ref{fig:exp}(c), we can experimentally realize a complete polarization-OAM hybrid-entangled BSM in a single photon, whereby all of the four Bell states, $\ket{\omega_{ca}^{\pm}}$ and $\ket{\xi_{ca}^{\pm}}$, are finally transmitted to four different ports, respectively.
This BSM scheme can be described as follows by four steps:
\begin{itemize}
\item[1.] The Mach-Zehnder interferometer and the QWP at $45^{\circ}$ (both head and tail are PBSs, and only one reflection difference between the two paths) are to make all four Bell states ($\ket{\omega_{ca}^{\pm}}$ and $\ket{\xi_{ca}^{\pm}}$) from non-separate states to separate states ($\ket{H}\ket{+\ell_0}$, $\ket{V}\ket{-\ell_0}$, $\ket{H}\ket{-\ell_0}$ and $\ket{V}\ket{+\ell_0}$).
\item[2.] Two of the four separate states are then projected through PBS and another two are reflected due to their polarizations.
\item[3.] The HWP, Sagnac interferometer, QWP and final PBS are made up of a OAM $\pm1$ sorter \cite{fu2015}.
\item[4.] Finally, SPPs ($+1$ or $-1$ in Fig.~\ref{fig:exp}) here are also to flatten photons' phase into a Gaussian mode that can be efficiently coupled into the SMF.
\end{itemize}
As mentioned above, all four Bell states can be discriminated in our experimental scheme, and there is no probabilistic photon loss in theory.

\begin{figure}[t]
	\centering
	\includegraphics[width=0.85\linewidth]{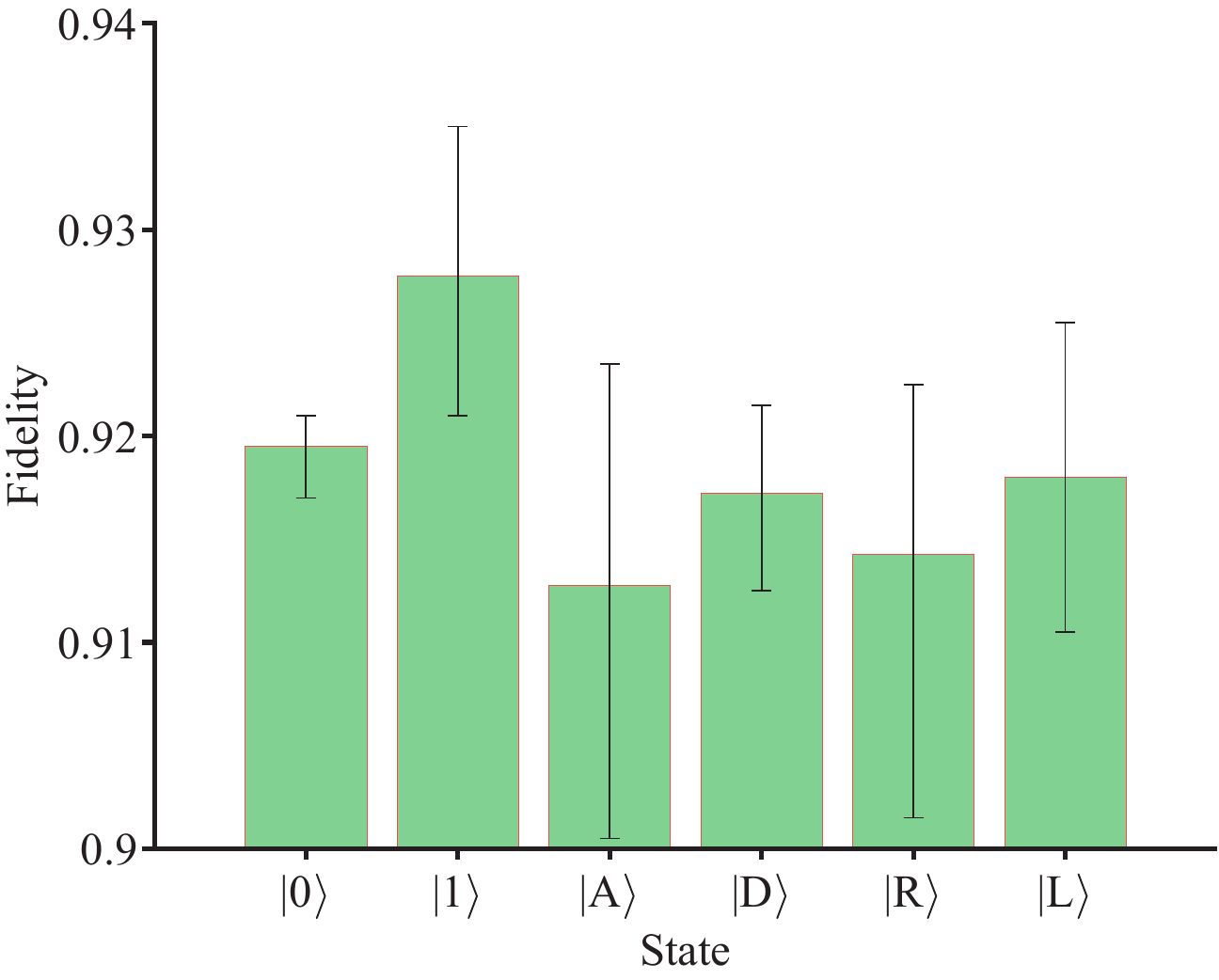}
	\caption{(color online) Fidelities of teleported states. The fidelity is calculated according to $F=\bra{\psi}\rho\ket{\psi}$, where $\ket{\psi}$ is one of the six mutually unbiased basis states. All fidelities are greater than the classical teleportation bound $F_{\mathrm{class}}=2/3$.}\label{fig:fidelity}
\end{figure}

In Sec. \ref{sec:proposal}, we have discussed that an alphabet of mutually unbiased states, such as the six poles of the Bloch sphere, can be used to test the present scheme based on quantum teleportation. We only need to check whether the fidelities for these input states are greater than $F_{\mathrm{class}}$ to ensure the use of quantum resources.
\begin{table}[t]
\caption{\label{tab:1}Ratio of counting probabilities of four ports (have been normalized).}
\begin{ruledtabular}
\begin{tabular}{ccccc}
Captures&A&B&C&D\\
\hline
Theory(\%)&25&25&25&25\\
Experiment(\%)&$23.8\pm1.3$&$25.4\pm1.2$&$24.6\pm0.8$&$26.2\pm1.1$\\
\end{tabular}
\end{ruledtabular}
\end{table}
FIG.~\ref{fig:Alltomo} shows the teleported states' tomography, which are reconstructed using a maximum likelihood estimator of the density matrix. FIG.~\ref{fig:fidelity} shows the fidelities for each of the six input states.
It can be seen that the six input states are well reproduced,
all these fidelities are greater than $F_{\mathrm{class}}$, indicating that it is actually a non-classical cloning experiment.
In addition, for all the data in the above measurements, we also have counted the ratio of coincidence events of the four ports, A, B, C and D (see in FIG.~\ref{fig:exp}).
The results are listed in TABLE~\ref{tab:1}, which is consistent with the theoretical probabilities (all being 0.25) for four Bell states.

\section{Conclusion and discussion}
\label{sec:conclusion_and_discussion}
In summary, we have experimentally demonstrated a process of quantum state transfer from one photon's polarization DoF to another photon's OAM DoF with a $91.8\pm1.3\%$ average fidelity.
We remind that although our transmitted state is known by Alice, this scheme has some similar advantages to original teleportation. One is that the transfer of materials bodies is not required. Another one is that only a part of information of the state is transferred via a classical channel. In other words, the state is protected from Eve.
In some sense, it could also be recognized as a kind of interface for the quantum state transfer between the different particles and different DoFs.
This interface offers a probability  that makes it to be employed in other hybrid quantum systems.
Meanwhile, we have verified experimentally that the hybrid-DoF BSM proposed in this paper can indeed achieve a higher Bell-efficiency, since all of the four hybrid-entangled Bell states can be discriminated well.
In addition, as far as the usage of OAM DoF one motivation is mainly inspired by its high-dimensional characteristic for the quantum information processing ($\ell=\pm1,\pm2,\cdots$). Although $\ell$ is limited to $\pm1$ values in this experiment, the value of $\ell$ can be actually set into any integers for each state by changing SPP in the entanglement source and in the BSM projection in principle.
Furthermore, our polarization-orbital coupled BSM scheme is entitled to be employed for OAM to polarization quantum state transfer and dense coding.

\begin{acknowledgments}
This work was in part supported by the National Nature Science Foundation of China (Grant Nos. 11534008, 11804271, 91736104 and 12074307), Ministry of Science and Technology of China (2016YFA0301404) and China Postdoctoral Science Foundation via Project No. 2020M673366.
\end{acknowledgments}

\nocite{*}

%

\end{document}